\RequirePackage{fix-cm}
\documentclass[twocolumn]{svjour3}          
\smartqed  
\usepackage{graphicx}
\usepackage{multirow}
\usepackage{amssymb}
\usepackage{amsmath}
\usepackage[hidelinks]{hyperref}
\usepackage[toc,page]{appendix}
\usepackage{placeins} 
\usepackage[binary-units=true]{siunitx}
\usepackage{xcolor}

\newcommand{\dunetpc}{\texttt{dunetpc}~}
\newcommand{\veight}{\texttt{v08\_24\_00}~}
\newcommand{\vnine}{\texttt{v09\_10\_00}~}

\journalname{Computing and Software for Big Science}

\begin{document}

\title{Deep Learning strategies for ProtoDUNE raw data denoising}

\author{Marco Rossi         \and
        Sofia Vallecorsa
}

\institute{Marco Rossi \and Sofia Vallecorsa \at
              CERN openlab, Geneva 23, CH-1211, Switzerland  \\
           \and
           Marco Rossi \at
              TIF Lab, Dipartimento di Fisica, Universit\`a degli Studi di Milano and INFN Sezione di Milano, Via Celoria 16, 20133, Milano, Italy \\
              \email{marco.rossi@cern.ch}
}

\date{Received: date / Accepted: date}

\maketitle

\begin{abstract}
In this work, we investigate different machine learning-based strategies for
denoising raw simulation data from the ProtoDUNE experiment. The ProtoDUNE detector
is hosted by CERN and it aims to test and calibrate the technologies for DUNE,
a forthcoming experiment in neutrino physics. The reconstruction workchain
consists of converting digital detector signals into physical high-level
quantities. We address the first step in reconstruction, namely raw data
denoising, leveraging deep learning algorithms. We design two architectures
based on graph neural networks, aiming to enhance the receptive field of basic
convolutional neural networks. We benchmark this approach against traditional
algorithms implemented by the DUNE collaboration. We test the capabilities of
graph neural network hardware accelerator setups to speed up training and
inference processes.
\keywords{
  Deep Learning \and ProtoDUNE \and Denoising \and Convolutional Neural
  Networks \and Graph Networks
}
\end{abstract}

\section{Introduction}
\label{sec:intro}
Deep learning algorithms achieved outstanding results in many research fields
in the last few years. The neutrino high energy physics community~\cite{domine2020,abi2020cnn,aurisano2016,kronmueller2019},
and in general the particle physics one~\cite{albertsson2019,bourilkov2019}, is making an effort
to apply such technologies to create a new generation of automated
tools. In general capable of processing
efficiently huge amounts of information, these algorithms work with
increased performance to mine deeper in collected data and discover hidden
patterns responsible for potential new physics scenarios. Event reconstruction
algorithms, i.e. reconstruction, the process of extraction of useful quantities from 
detector or simulated data, are especially suited to the application of this approach.

DUNE~\cite{abi2020tdrI} is a next-generation experiment in the neutrino
oscillation research field. The DUNE Far Detector (FD)~\cite{abi2020tdrII,abi2020tdrIII,abi2020tdrIV},
based at the Sanford Underground Research Facility (SURF) in South Dakota, will
be the largest monolithic Liquid Argon Time Projecting \linebreak
Chamber (LArTPC) detector ever built. In order to test and validate technologies
for the construction of such detector, a prototype, ProtoDUNE Single Phase
(SP)~\cite{abi2017pdunetdr}, has been built at the CERN Neutrino Platform.

Particle interactions with the liquid Argon produce ionization electrons that are drifted,
thanks to an electric field, towards Anode Plane Assemblies (APAs) \linebreak made of three
readout planes that collect the deposited charge through time. Each readout plane is
a bundle of wires, oriented in a specific direction and continuously monitored by
the detector: two planes are called induction planes, while the last one is named 
the collection plane. In particular, the ProtoDUNE SP cage envelope is surrounded by 6
different APAs, each with two induction planes of 800 wires and a collection one,
holding together 960 wires. Overall, a total of 15360 wires on the sides of the
detector provide a comprehensive and detailed picture of events from different
points of view.

ProtoDUNE SP measures a digitized value of the induced current on each wire (ADC
value). Since the detector maximum sampling rate is $\SI{2}{\mega\hertz}$ and the
common readout time windows at ProtoDUNE SP last \linebreak $\SI{500}{\nano\second}$, raw
data (or raw digits) form a sequence of $6000$ current measurements per wire. In
this paper we focus on event reconstruction of ProtoDUNE SP simulated data. We
simulate interactions with the help of the LArSoft~\cite{church2014larsoft}
framework and its \dunetpc package. Figure~\ref{fig:data v8}, shows an example
of the simulated data from a single collection plane: raw digits are cast into
a $6000\times960$ resolution image
plotting the ADC values heat map over time versus wire number axes.

Raw digits recorded by detector electronics intrinsically contain noise that must
be filtered out. The traditional approach~\cite{abi2020tdrII},
developed by the MicroBooNE experiment~\cite{acciarri2017,adams2018},
is based on a 2-dimensional deconvolution and consists of two steps: first, a mask
is produced to identify the Regions Of Interest (ROI) in the raw data containing
signals; then, in those regions Gaussian shape peaks are fitted to match the inputs,
filtering them in Fourier space and deconvolving back the results.

The goal of this work is to tackle the denoising problem, implementing automatic
tools at the readout plane view level. The nature of the inputs, above all
sparsity and size, represents a challenging benchmark for Deep Learning models.
Images indeed contain signals, mainly organized in long monodimensional tracks and
clusters, divided by almost empty extended regions. Then, trying to train
classic deep feed-forward models on sparse inputs might lead to sub-optimal
results, since gradients would receive a large contribution from pixels in those
noise-dominated empty regions. Moreover, the high image resolution might result
in a pixel receptive field limited to just local neighborhoods, which are small
compared to the extension of the meaningful spatial features included in the
plane views. Finally, handling the size of the inputs requires careful RAM and
GPU memory management while dealing with complex model architectures.

The paper is organized as follows. First, in section~\ref{sec:propmod} we describe the
models and operations used to tackle the problem. Then, section~\ref{sec:dattrain}
illustrates the dataset we generated and the methods employed to train the
networks. Section~\ref{sec:exp} is devoted to the presentation of
experimental results. Finally, in section~\ref{sec:concl} we present our 
conclusion and future development directions.

\begin{figure}
  \label{fig:gconv}
  \centering
  \includegraphics[scale=0.1]{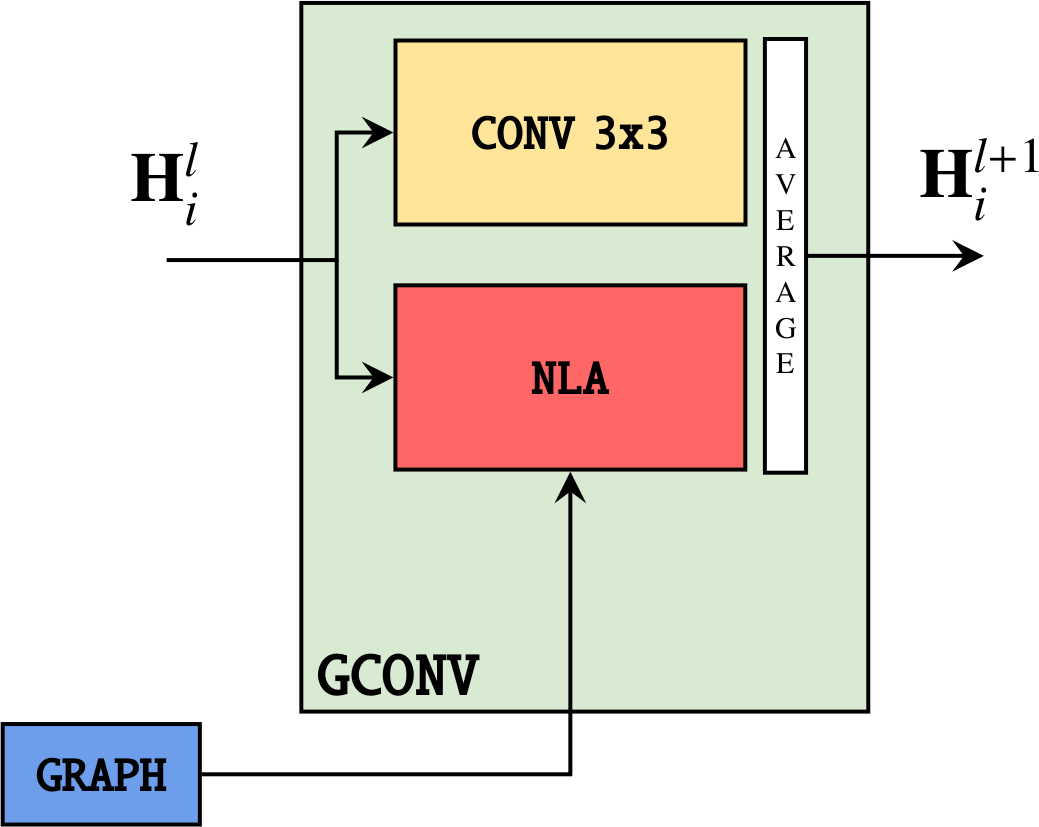}
  \caption{GCONV layer. The inner representation vector $\mathbf{H}^{l}_i$ is updated
  to $\mathbf{H}^{l+1}_i$ by means of NLA and 2D convolution operations. NLA
  relies on a previously computed KNN graph.}
\end{figure}

\section{Proposed Models}
\label{sec:propmod}

\begin{figure*}
  \caption{GCNN architecture. The model takes noisy input images $\texttt{x}$ and
  outputs denoised ones \texttt{\textbf{x}$_{\mathrm{DN}}$}. The design is organized with Low and High Pass
  Filters (LPF, HPF) as in~\cite{valsesia2019deep}. As explained in
  section~\ref{subsec:train}, we concatenate a pretrained ROI block with
  preprocessing layers to make the network distinguish between signal and
  background. The ROI block, indeed, performs the binary segmentation. }
	\label{fig:gcnn}
  \includegraphics[width=\textwidth]{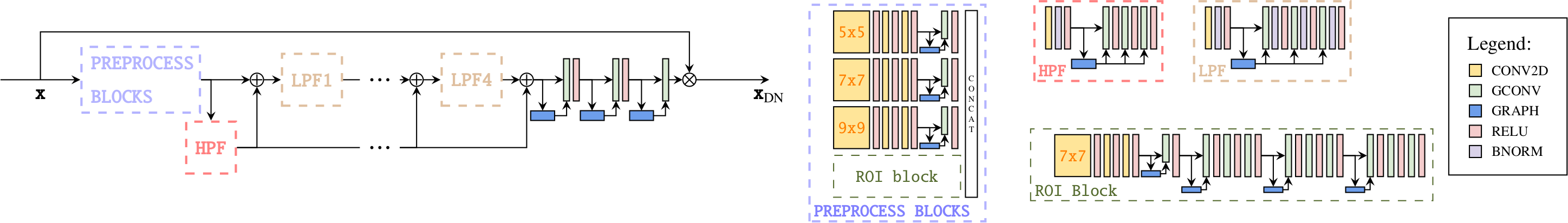}
\end{figure*}

Convolutional neural networks (CNN) are based on a stack of sliding kernels comprised
of multiple filters, that are trained according to some optimization method to
output a feature map. The convolutional kernel generates each feature map pixel
as a function of a small neighboring portion of the input image. Hence, the
receptive field of the pixels in each layer is constrained by the kernel size,
which could be enlarged by increasing the depth of the network. 
The scope of the present section is to describe an alternative approach to the
issue: by enriching the network with alternative operations we hope to increase the
expressiveness of the internal representation by exploiting non-local correlations
between pixel values, alongside the already discussed local neighborhood pixel intensities.

\label{subsec:gcnn}
\subsection{Graph Convolutional Neural Network}

We implement a Graph Convolutional Neural Network (GraphCNN) inspired
by~\cite{valsesia2019image,valsesia2019deep} and based on the Edge Conditioned
Convolution (ECC) operation, first presented in~\cite{simonovsky2017dynamic}. We
employ a simplified version of the ECC layer: it builds the output
representation as a pixel-wise average of a common convolution with a $3\times3$
kernel and a Non-Local Aggregation (NLA) operation. We wrap such operations into
a network layer called Graph Convolution (GCONV). Figure~\ref{fig:gconv} sketches
the GCONV layer mechanism.

The NLA connects each pixel to its $k$ closest ones in feature space, according
to the Euclidean distance, and mixes the information through a feed-forward layer.
If at layer $l$, the $i$-th of an $n$-pixel input image is described by the vector
$\mathbf{H}^l_i \in \mathbb{R}^{d_l}$, then the NLA output $\mathbf{H}^{l+1}_i$
has the following form:
\begin{equation}
  \scriptsize
  \mathbf{H}^{l+1}_i = \sigma \Biggl( \frac{1}{|\mathcal{N}_i^l|}
                       \sum_{j\in \mathcal{N}_i^l} 
                        \mathrm{\Theta}^{l} \bigl(\mathbf{H}^l_i-\mathbf{H}^l_j \bigr)                           
                      + \mathrm{W}^l \mathbf{H}^l_i + \mathbf{b}^l
  \Biggr) \hfill \in \mathbb{R}^{d_{l+1}}
  \label{eq:nla}
\end{equation}
With $\mathcal{N}_i^l$ being the neighborhood of pixel $i$ in the $H_i^l$
representation at layer $l$;
$\{\mathrm{\Theta}^{l} , W^l\} \in \mathbb{R}^{d_{l+1}\times d_l}$ and
$\mathbf{b}^l \in \mathbb{R}^{d_l+1}$ are trainable weights and biases shared
throughout pixels; $\sigma$ is the element-wise sigmoid function.

Note that the operation in equation~\ref{eq:nla} requires building a k-NN graph
which requires an amount of memory proportional to the area of the
input image times the number of neighbors per pixel $k$. Assuming we fix $k=8$,
with an input image of $960\times6000$ pixels and employing single precision
floating point numbers, the graph construction operation burden is of order
$\mathcal{O(\SI{200}{\mega\byte})}$. If the architecture involves multiple graph
building operations, the GPU memory is easily saturated. Therefore, we
have to limit the model inputs to just crops of the actual image as explained in
section~\ref{subsec:train}.

Figure~\ref{fig:gcnn} shows our GCNN network architecture. Note the final
residual connection in the top branch: the usual sum has been replaced by a
multiplication. This choice is tailor made on the input data themselves, which
are mainly comprised of long tracks separated by empty space. In those regions
it could be easier to learn how to remove the noise multiplicatively rather than
additively. The network, in principle, doesn't have to learn to perfectly
profile the noise and then subtract it from the input itself, whereas it can
employ a mask to cut down such uninteresting regions with multiplications by
small numbers.

\subsection{U-shaped Self Constructing Graph Network}
\label{subsec:uscgn}

\begin{figure*}
  \caption{USCG-Net architecture. The Pooling block has a two folded aim thanks
  to its adaptive pooling layer: on the left branch it downscales the input,
  while on the right one it provides upsampling. According to~\cite{xie2017aggregated},
  ResNeXt-50 with $32\times4$ template is arranged in 4 main blocks, containing
  9, 12, 18 and 9 convolutional layers respectively. In the chart we refer to such blocks as
  \texttt{ResNeXt-50<block number>} and we insert a residual connection
  between the second and the third block. The $1\times 1$ convolutions in the
  horizontal links are needed to adapt the number of filters in the image to
  perform the residual recombination operation.}
  \label{fig:uscgn}
  \centering
  \includegraphics[width=0.7\textwidth]{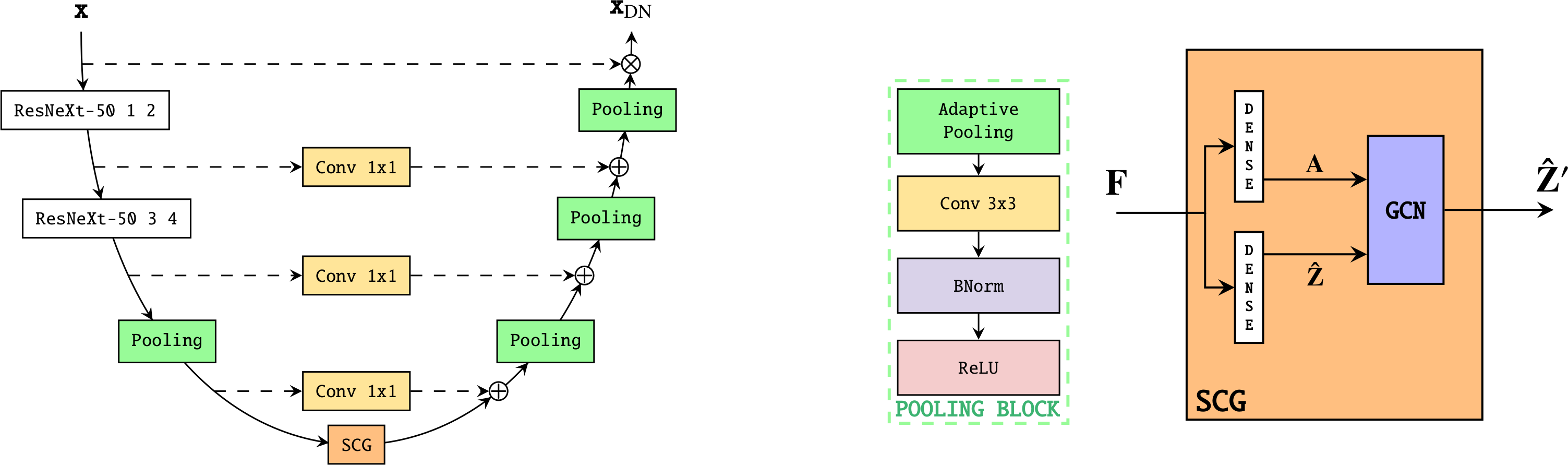}
\end{figure*}

As stated in the previous section, the main limitation of the GCNN model is the
memory consuption burden due to the graph operation: graph
building scales linearly with the size of the inputs, therefore, the model
accepts crops rather than the full APA image. The cropping workaround prevents
very long range correlations between pixels being taken into account and leads
to inference time performance issues. If the user does not have access to a high
GPU memory bandwidth, only batches of crops can be processed in parallel and
processing big datasets is not time efficient.

As an alternative approach to the GCNN, we follow the idea introduced in~\cite{liu2020scgnet} with the Self
Constructing Graph Network (SCG-Net): a graph neural network that outputs results
from a low dimensional representation of the high resolution image created by a
full CNN; the original shape of the image is then retrieved interpolating between pixels.
In the original work, a bilinear interpolation is employed for the upsampling.
In this way the authors were able to process big images containing up to
$6000\times6000$ pixels. This seems a more natural approach to the problem when the
input images contain dense features, rather than sparse and localized ones as
in our case. We believe that the entire pipeline potentially washes out the fine
grained information contained in the input during downscaling, which cannot be
recovered by means of a simple interpolation.

Therefore we introduce the U-shaped Self \linebreak Constructing Graph Network (USCG-Net), where
a U-Net~\cite{ronneberger2015unet} like network structure with residual connections
carries the information from the input node all the way to the outputs.
Figure~\ref{fig:uscgn} shows the USCG-Net architecture, with pooling blocks,
comprised of convolutional and pooling layers, that take care of stepwise scaling the inputs.
A pretrained ResNeXt-50 with a $32\times4$ template~\cite{xie2017aggregated} is used to build an
initial feature map to be fed into the SCG layer. The early-layer representation
of the pretrained network should be generic enough to catch the spatial features of
the inputs, driving the training process during the initial phases of the optimization.
Note the residual recombination operations in the right branch: the sum in the
residual connections has been replaced by a convolution with a $1\times 1$
kernel in order to increase the complexity of the network. Finally, for the
final residual link, we employ again the multiplication trick, as explained in
section~\ref{subsec:gcnn}.

The SGC layer is the core of the network: the input image is turned into a graph,
allowing connections between distant pixels. The SCG layer~\cite{liu2020scgnet}
takes in an image $\mathbf{F} \in \mathbb{R}^{h\times w\times d}$ and
through an encoding functions, represented by dense layers,
maps it into two outputs: an adjacency matrix $\mathbf{A} \in \mathbb{R}^{n\times n}$,
describing the graph edges, and a node feature vector $\mathbf{\hat{Z}} \in \mathbb{R}^{n\times c}$.
In the above formulas, $d$ and $c$ are the input and output channel dimensions
respectively and $n$, which is equal to the input image resolution $h\times w$,
is the number of extracted nodes. The predicted graph is then analyzed by
a 1-layer graph neural network, such as GCN~\cite{scarselli:2009} or
GIN~\cite{xu2019powerful}, to further mix the node features, yielding a final node
feature vector $\mathbf{\hat{Z}'} \in \mathbb{R}^{n\times c'}$. In our experiments
we exploited the GCN layer as default; architecture variants with other kinds of
graph layers can be addressed in a future work. $\mathbf{\hat{Z}'}$
can be finally projected back into $\mathbb{R}^{h \times w \times c'}$ and
upsampled by pooling blocks in the right branch of the USCG-Net to the original
image resolution.


\section{Dataset and Training}
\label{sec:dattrain}

\subsection{Datsets}
\label{subsec:datasets}

\begin{table}
  \caption{Datasets for training and testing. The two differ in the producer
  package version, the size and the event beam energies. The second dataset contains
  $10$ events for each proton energy specified.}
  \label{tab:datasets}
  \begin{tabular}{lll}
    \hline\noalign{\smallskip}
    \dunetpc & $n$ events & $p$ energy \\
    \noalign{\smallskip}\hline\noalign{\smallskip}
    \veight & 10 & $\SI{2}{\GeV}$ \\
    \multirow{2}*{\vnine} & \multirow{2}*{70} & $\SI{0.3}{\GeV}$, $\SI{0.5}{\GeV}$, $\SI{1}{\GeV}$, $\SI{2}{\GeV}$, \\
                          &    & $\SI{3}{\GeV}$, $\SI{6}{\GeV}$, $\SI{7}{\GeV}$\\
    \noalign{\smallskip}\hline
  \end{tabular}
\end{table}

\begin{figure*}
  \centering
      \includegraphics[width=0.32\textwidth]{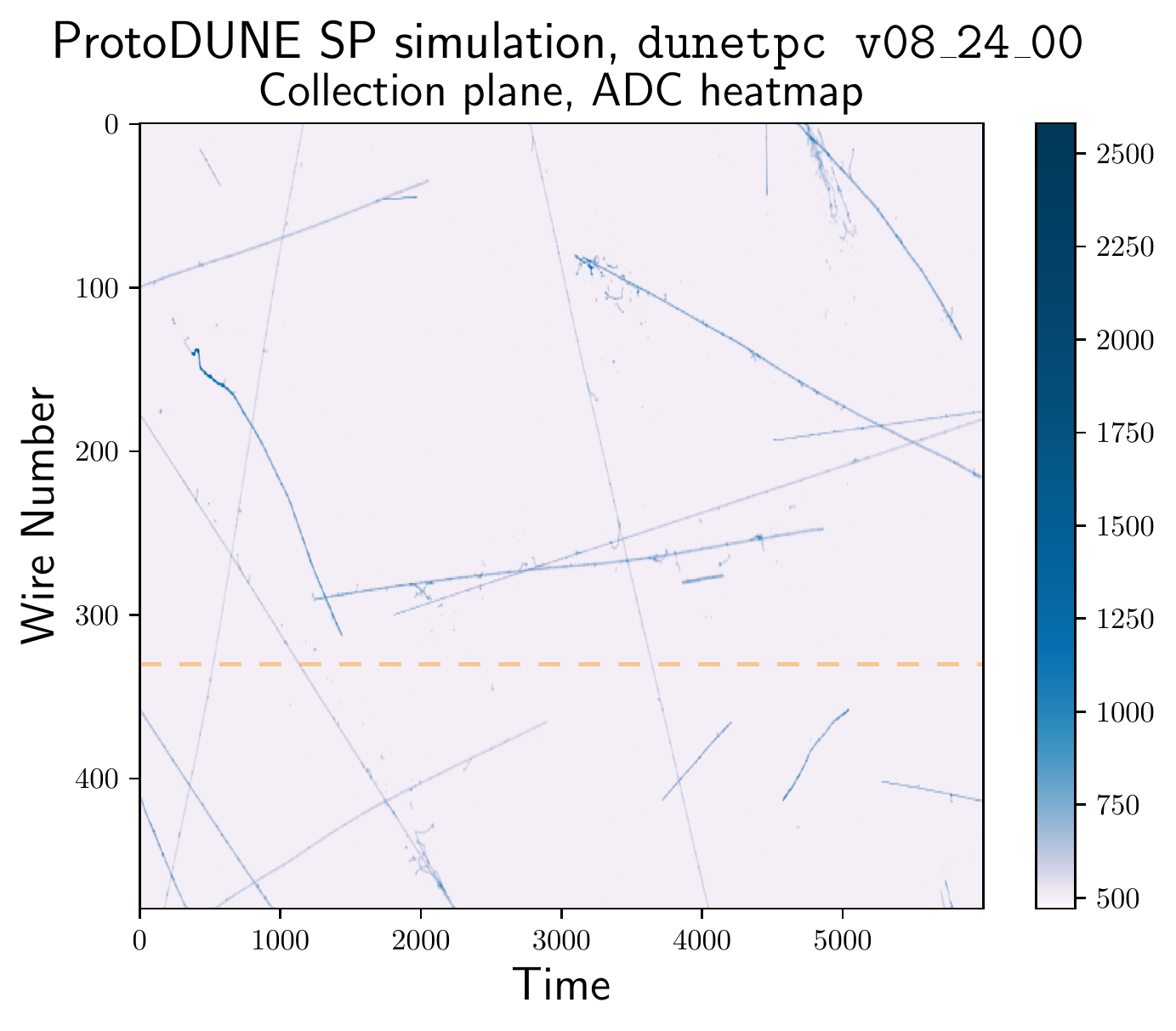}
      \includegraphics[width=0.32\textwidth]{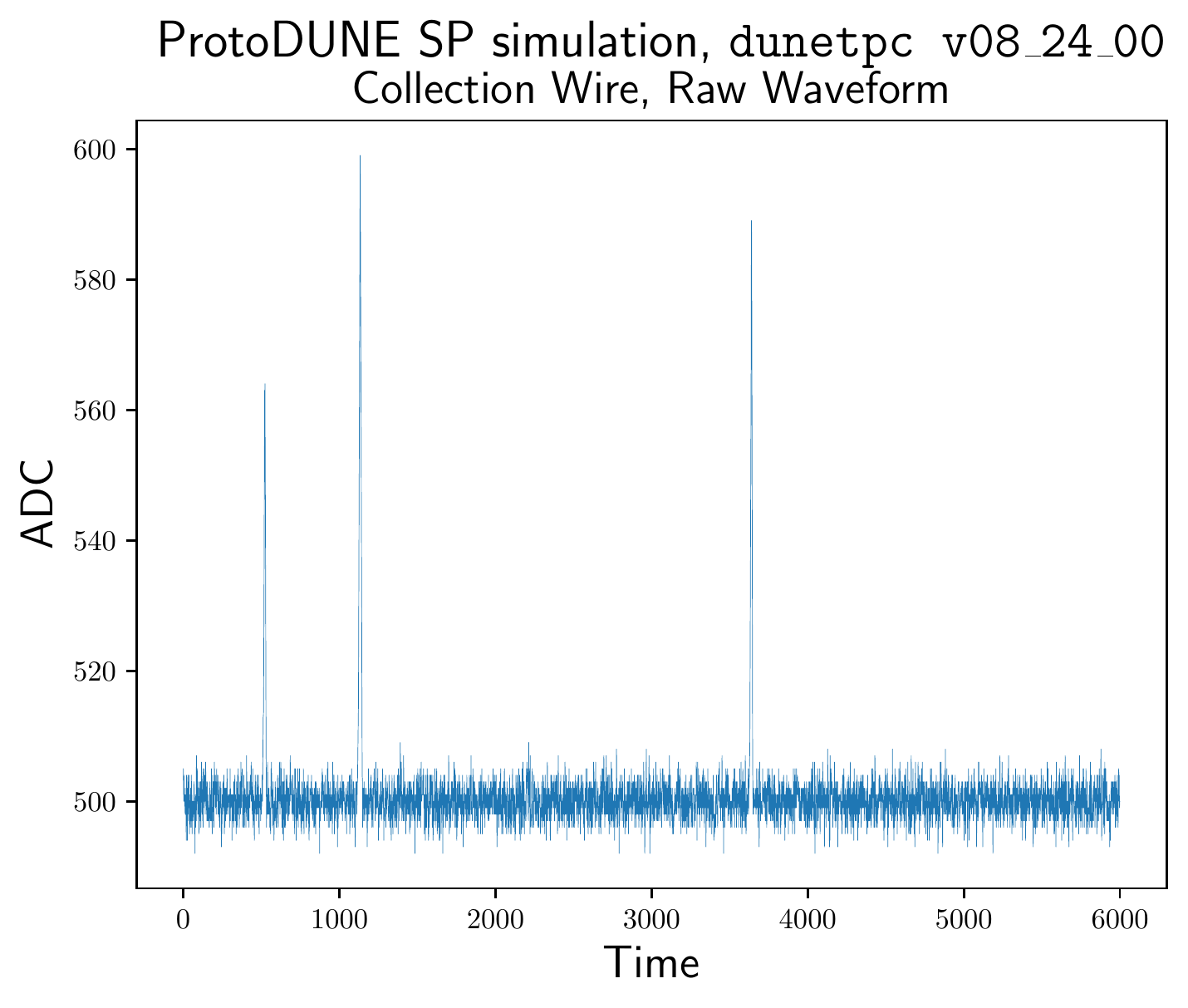}
      \includegraphics[width=0.32\textwidth]{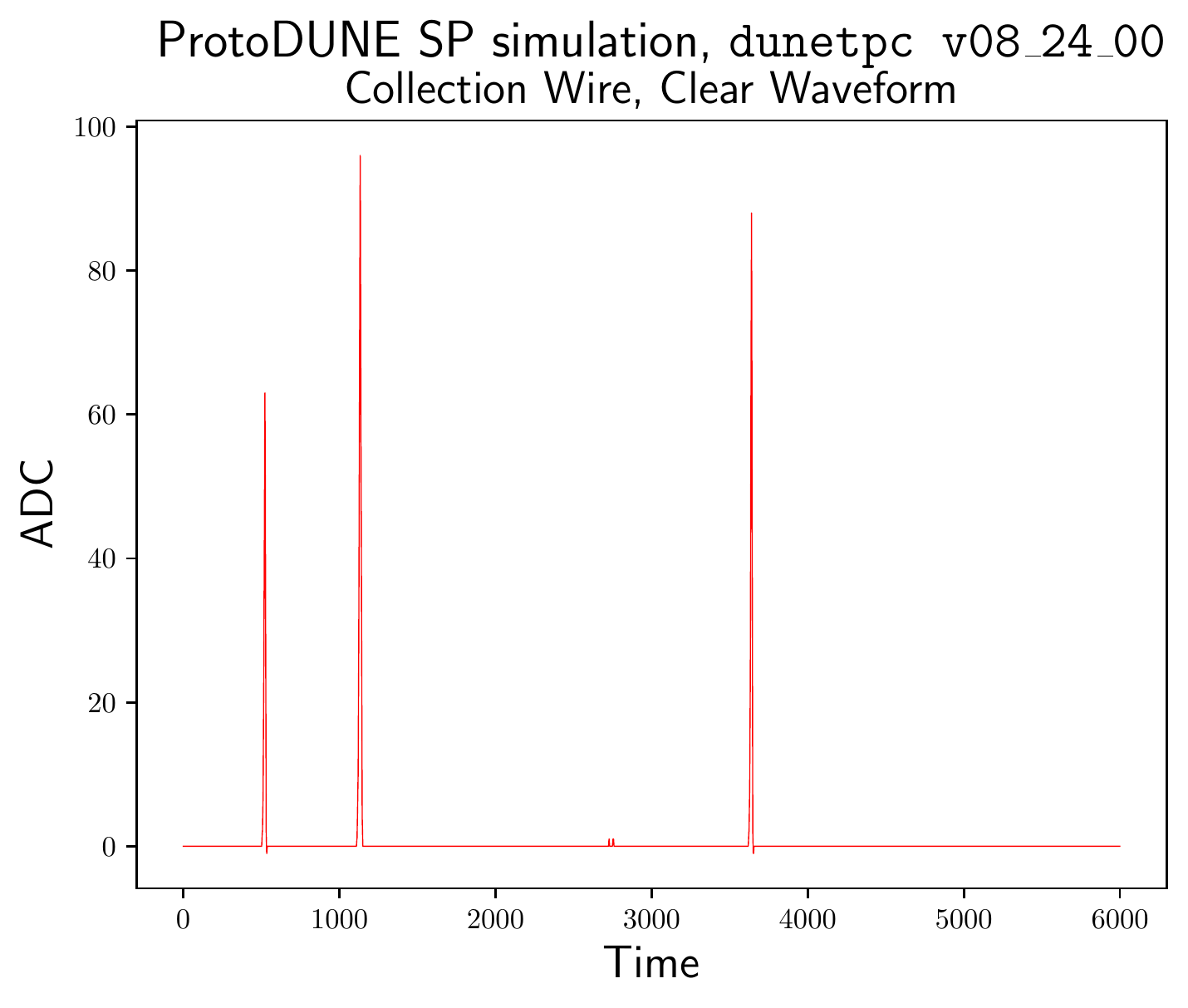}
  \caption{Example taken from \dunetpc \veight dataset. Collection plane view
  with noisy and clear waveforms extracted from the channel
  marked by the horizontal orange dashed line.}
  \label{fig:data v8}
\end{figure*}

\begin{figure*}
  \centering
      \includegraphics[width=0.32\textwidth]{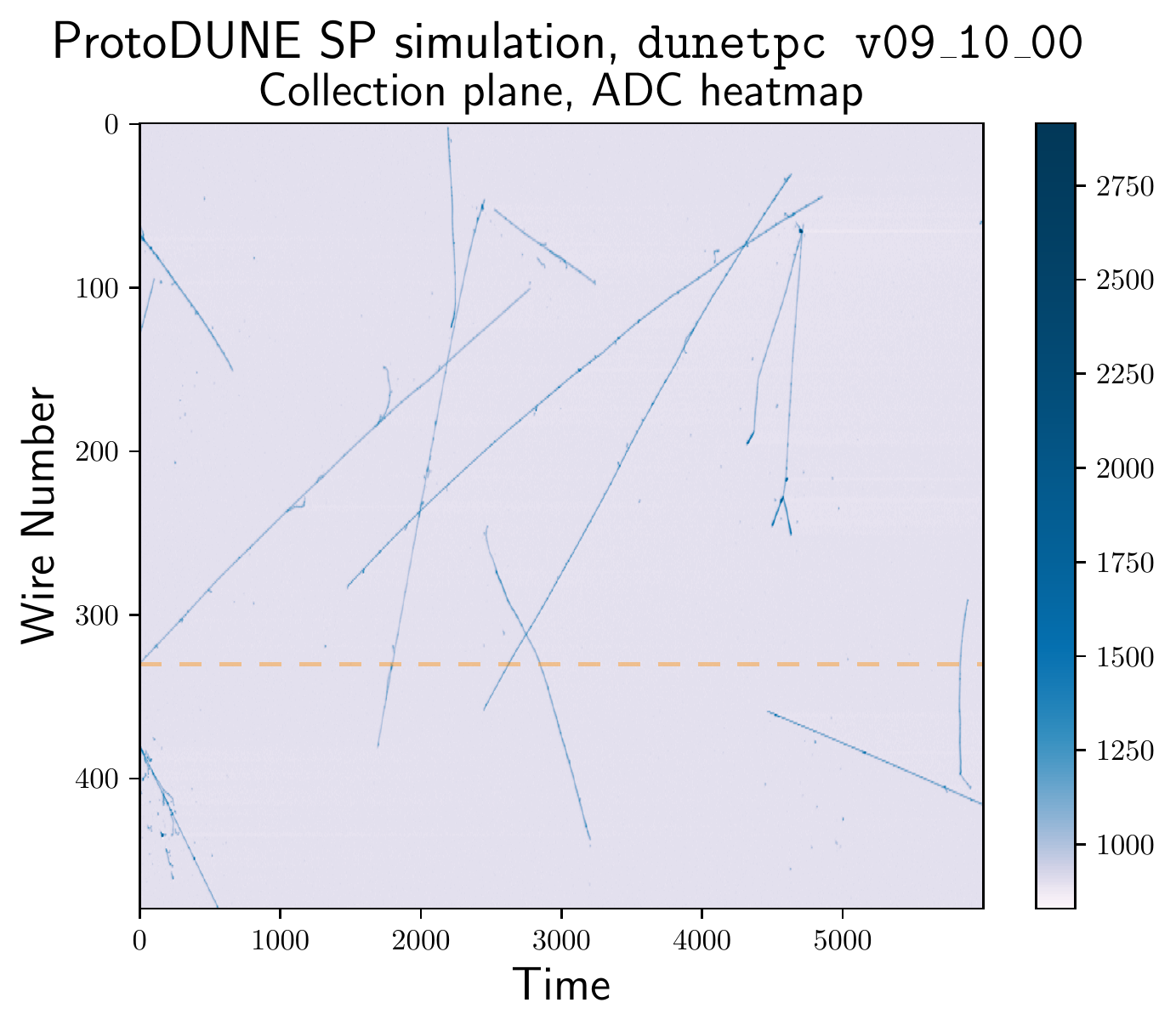}
      \includegraphics[width=0.32\textwidth]{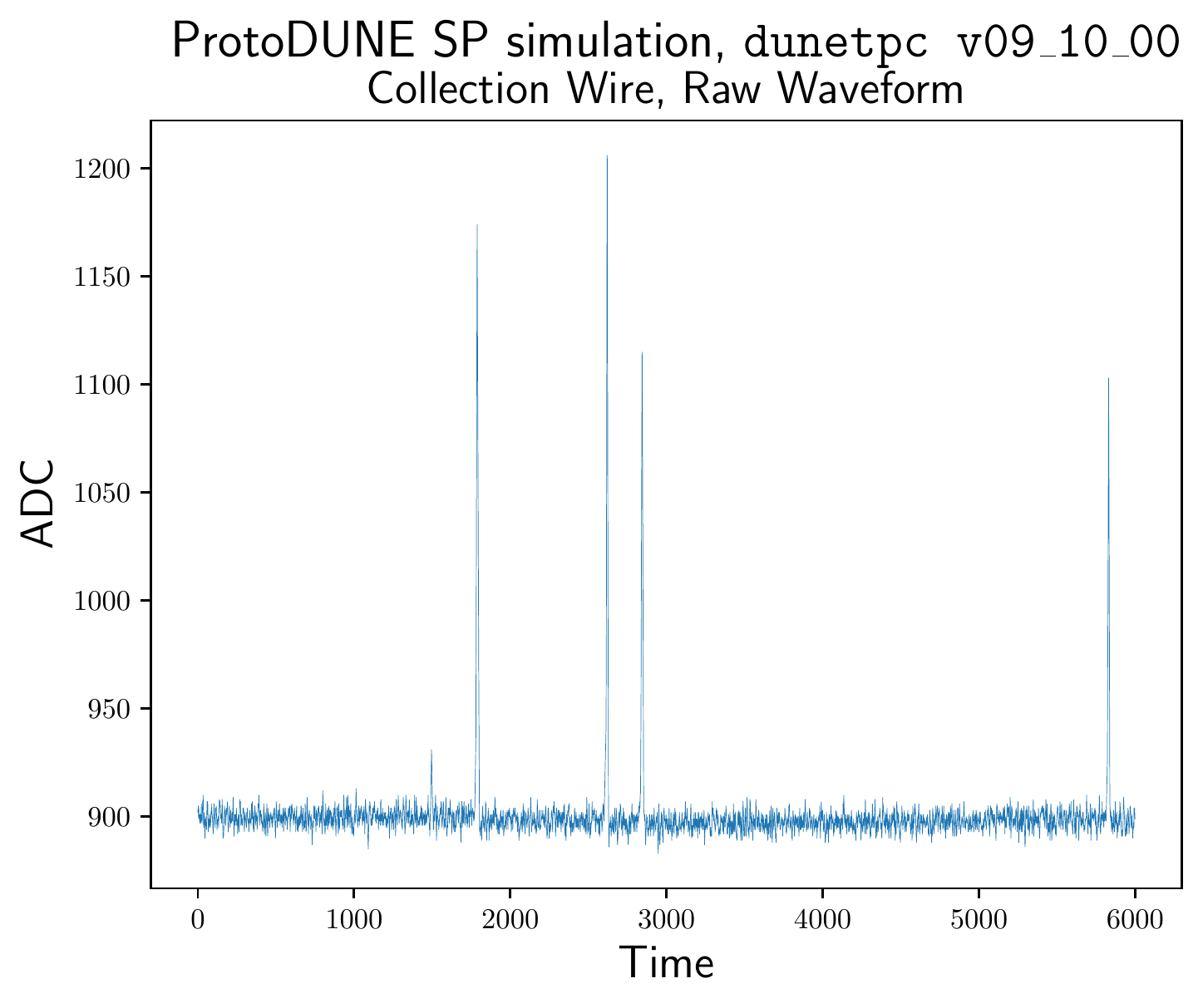}
      \includegraphics[width=0.32\textwidth]{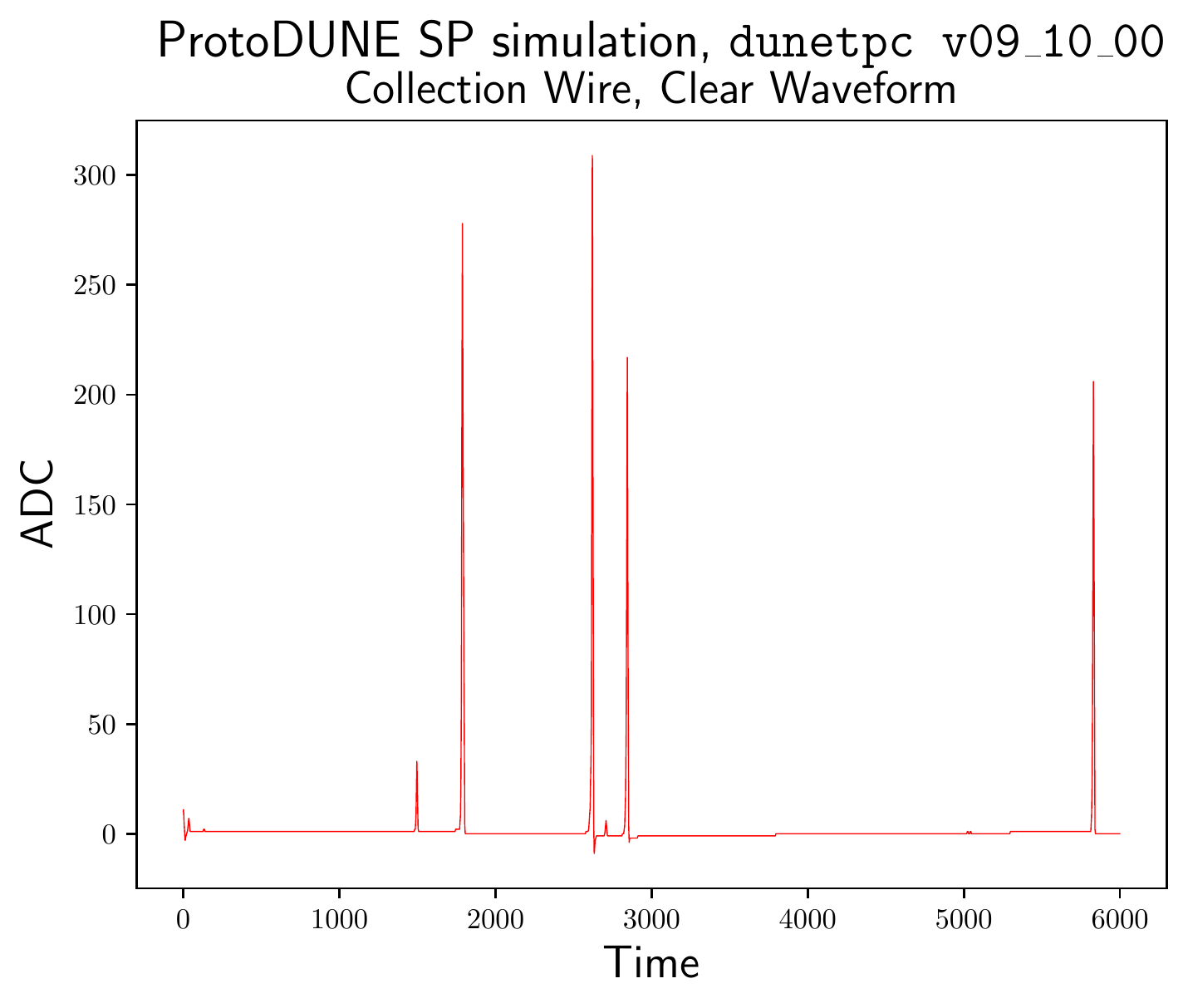}
  \caption{Example taken from \dunetpc \vnine dataset. Collection plane view
  with noisy and clear waveforms extracted from the channel
  marked by the horizontal orange dashed line.}
  \label{fig:data v9}
\end{figure*}

In this section we present the datasets used to train and test our models.
We underline that we present results for simulated data only, testing on detector
data is out of the scope of the present work. We simulate interactions within the ProtoDUNE
SP chamber through the LArSoft~\cite{church2014larsoft} framework and its \dunetpc package.
We consider events originating from a proton beam of various energies, plus cosmic
rays, interacting with the Argon targets. Our supervised training approach is based
on simulated raw digits. The simulation provides also the associated noise free charge
depositions: we employ this information as target outputs for the denoising models.

Table~\ref{tab:datasets} describes the considered datasets. In
figures~\ref{fig:data v8} and~\ref{fig:data v9}
we show visual examples from \veight \linebreak and \vnine datasets. We plot raw digit
images and horizontal slices, namely single channel waveforms, either with (raw
waveform) and without noise (clear \linebreak waveform). Noise is mainly comprised of a
pedestal value that changes across different datasets and a background
noise that overwhelms small energy depositions and modifies spikes amplitudes.
We consider the \linebreak \veight
dataset a simplified dataset since it is smaller and contains easier features to
denoise and segment than the \vnine one. \vnine is more complex due to the
detector data driven approach upon which the generator software is based. Clear
waveforms are not simply zero in the empty regions, but rather
contain small steps. Moreover, low frequency negative tails after big spikes
in clear waveforms are clearly visible in figure~\ref{fig:out v9}.

We hold out from our datasets $10\%$ of images for validation and $10\%$ for testing.
The validation set is used to choose the best model, while the test set is
employed to present the final results given in section~\ref{sec:exp}. The
criterion chosen to present results is to aggregate each performance quantity,
over the test or validation set, then report it as a symmetric
interval with the central value and its uncertainty being the arithmetic mean and
the standard deviation on the mean respectively. This choice also influences the best
model selection during training, namely, the network checkpoint at epoch end that
achieves the lowest upper bound of the loss function interval.

We remark that, as described in section~\ref{sec:intro}, a single event at ProtoDUNE SP is
recorded by 6 different APAs, that in turn provide 3 plane views. So one event
does not count as one training point for our neural networks. In
section~\ref{sec:propmod}, we anticipated that the GCNN network acts
indipendently on small crops: if $32\times32$ crops are employed, each event in
the dataset becomes $9\times10^4$ crops. The USCG-Net, instead, processes
entire plane views. The two datasets provide $18$ and $126$ plane views to assess
the model performance and its associated uncertainty. Note that the output image
has a dimensionality of the order of millions of pixels and each pixel value provides a
contribution in the computation of the final performance metric.

\subsection{Network Training}
\label{subsec:train}
All networks that we train share the same pre-process\-ing procedure, where we take the
plane views in a data\-set and apply a median subtraction to each of them. This
step is fundamental to estimate and subtract the pedestal value. We introduce
this operation mimicking the traditional approach because it contributes to improve
the final performance. Furthermore, for network training stability, the inputs are
rescaled in the unit range according to the min-max normalization technique.

The GCNN network suffers from the memory issue discussed in section~\ref{subsec:gcnn}.
To address this problem, we employ a data parallel approach, cropping the inputs
into $32\times 32$ pixels images and processing every tile independently.
However, this solution requires some subtleties in the training process due to
the nature of the inputs.

Since the inputs contain sparse features, it is likely that the most of the crops
contain little to no signal. It is meaningless to feed the the network with multiple
empty crops. In order to speed up the training, we decide to train on just a subset of the
available crops. This choice, in turn, triggers a second issue: sampling randomly
the subset of crops provides an extremely unbalanced dataset, where we are very
likely to miss crops containing interesting charge depositions.
Hence, we fix the percentage of crops containing signal to balance the signal to
background pixel ratio: in our experiments, we keep this
quantity at $99\%$.
The cropping procedure works as follows: we mark as signal all the pixels in 
clear waveforms that have non-zero ADC value count in a plane view; we sample
randomly the desired number of crop centers, complying with the exact ratio of
signal to background percentage; we finally crop the plane views around the
drawn centers.

We train the GCNN network in figure~\ref{fig:gcnn} for both the image segmentation task 
and denoising, managed by the ROI Block and the network final outputs
respectively. Image segmentation is a classification task at the pixel level,
where the labels usually correspond to a finite set of exhaustive and mutually
exclusive classes: the ROI block, indeed, is trained to distinguish between pixels
that contain or not electron induced current signals. Image denoising,
instead, aims at subtracting the noise contribution from each pixel intensity and
is a regression task.

In order to optimize the two parts of the network, we
design an ad-hoc training strategy as follows. First, we train the ROI Block
alone on image segmentation for $100$ epochs and save the best configuration: in
this step the network branch learns to distinguish between signal and background,
i.e. empty pixels. We employ binary cross-entropy as the loss function,
along with the AMSGrad variant~\cite{j.2018on} of the Adam algorithm~\cite{kingma2017adam}
as the optimizer with a learning rate of $10^{-3}$. Second, we freeze
and attach the trained ROI block weights to the remaining part of the network.
Then, we train the GCNN on image denoising for a further $50$ epochs and save the
network's best configuration.

The GCNN denoising model is trained using the AMSGrad optimizer
with a learning rate of $9\cdot 10^{-3}$, while optimizing a custom loss function made
of two contributions: the mean squared error (MSE) $\mathcal{L}_{MSE}$ between
labels and outputs and a loss function $\mathcal{L}_{ssim}$ derived from the
statistical structural similarity index measure (stat-SSIM)~\cite{chann:2008}.
stat-SSIM for two images $\mathbf{x}$ and $\mathbf{y}$, is given by the
following equation:
\begin{equation}
  \label{statssim}
  \begin{aligned}
  \mathrm{stat{\text-}SSIM} (\mathbf{x}, \mathbf{y}) &= \frac{1}{n_p n_c}\sum_{pc}\bigg( \frac{2\mu_x\mu_y + \epsilon_\mu}{\mu_x^2 + \mu_y^2 + \epsilon_\mu} \bigg)_{pc}\\
  \times & \bigg( \frac{2\mathrm{E}[(\mathbf{x}-\mu_x)(\mathbf{y}-\mu_y)] + \epsilon_{\sigma}}{\mathrm{E}[(\mathbf{x}-\mu_x)^2] + \mathrm{E}[(\mathbf{y}-\mu_y)^2] + \epsilon_{\sigma}} \bigg)_{pc}
  \end{aligned}
\end{equation}
where $\mu_i$ is a shorthand for the image $\mathbf{i}$ expected value $E[\mathbf{i}]$, which
is computed through a convolution with a $11 \times 11$
Gaussian kernel of standard deviation $3$. All the other expectation values in
the equation are calculated convoluting the argument quantity with the same Gaussian filter.
$\epsilon_\mu$ and $\epsilon_\sigma$ are two regulators that limit the maximum resolution
at which the fractions in the equations are computed, respectively imposing a cut-off on the
mean and variance expected values. In particular, when both the numerator and the denominator
of a fraction reach much smaller values than the corresponding $\epsilon$, the output gets close to one.
In the experiments, $\epsilon_\mu$ and $\epsilon_\sigma$ are fixed to the square
of 0.5. This choice implies that for the stat-SSIM computation, we estimate
the means and standard deviations of the distributions at scales larger than half of one ADC value,
namely the granularity of the recorded detector hits.
The result is finally averaged over the entire image containing $n_p$ pixels and $n_c$ channel dimensions.
The quantity in equation~\ref{statssim} takes values in
the range $[-1,1]$ and approaches $1$ only if $\mathbf{x}=\mathbf{y}$. The
associated loss function is then given by $\mathcal{L}_{ssim} = 1 - \mathrm{stat{\text-}SSIM}$:
it is a perceptual loss, in the sense that tries to assess the fidelity of the image focusing
on structural information. It relies on the idea that pixels may have strong
correlations, especially when they are spatially close. In contrast, MSE
evaluates pixels' absolute differences, without taking into account any
dependence amongst them. More details on the interpretation of these quantities
can be found in~\cite{wang:2009mse}. The two contributions in the loss function are weighted as follows:
$\mathcal{L} = \alpha \cdot \mathcal{L}_{MSE} + (1-\alpha) \cdot w \cdot \mathcal{L}_{ssim}$.
We fine tune the multiplicative parameter $w=10^{-3}$, to balance the gradients
w.r.t. the model's trainable parameters provided by the two terms in the sum. The
parameter $\alpha$ is fixed to $0.84$ as in~\cite{zhao2018loss}.

In the following we will refer, with slight abuse of notation, to a CNN as a GCNN network
with Graph Convolutional layers replaced by plain Convolutional ones.

The USCG-Net is relatively simple to train: since no crops are employed, no
sampling method is required. Although no cropping is needed, since the model fits in a
single $\SI{16}{\giga\byte}$ GPU even with an entire plane view as input, we prefer
to employ a sliding window mechanism as in the original SCG-Net paper. We split
the raw digits array along the time axis with a 2000 pixel wide window and 1000 
pixel stride and feed each slice to the network. The results are then combined
by averaging predictions on overlapping regions. The USCG-Net is trained
minimizing the MSE function between model outputs and clear waveforms, with
AMSGrad optimizer and learning rate of $10^{-3}$. In this case, we drop the
stat-SSIM contribution from the loss function, since we experienced training
convergence problems when including that term during our experiments.

\section{Experiments results}
\label{sec:exp}
We employ four different metrics to assess the goodness of our models and benchmark them against
the state-of-the-art approach. Three of them are the stat-SSIM, peak signal to noise ratio
(PSNR) and MSE on the two dimensional raw digits, while the last one is a custom metric called
integrated mean absolute error (iMAE) and will be defined in the following.

\begin{figure*}
  \centering
      \includegraphics[width=\textwidth]{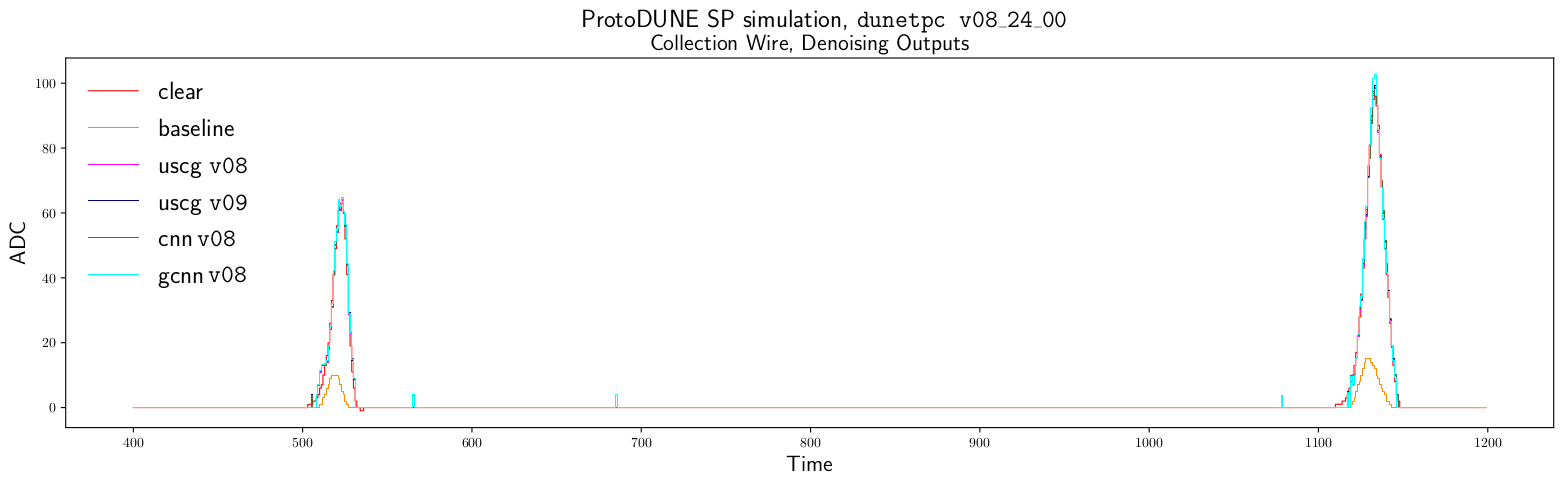}
  \caption{Detail of a raw waveform from  \dunetpc \veight dataset: label,
  traditional algorithm and neural networks outputs. The version,
  \texttt{v08} or \texttt{v09}, next to the model name in the legend
  refers to which dataset the correspondent model was trained on.}
  \label{fig:out v8}
\end{figure*}

\begin{figure*}
  \centering
      \includegraphics[width=\textwidth]{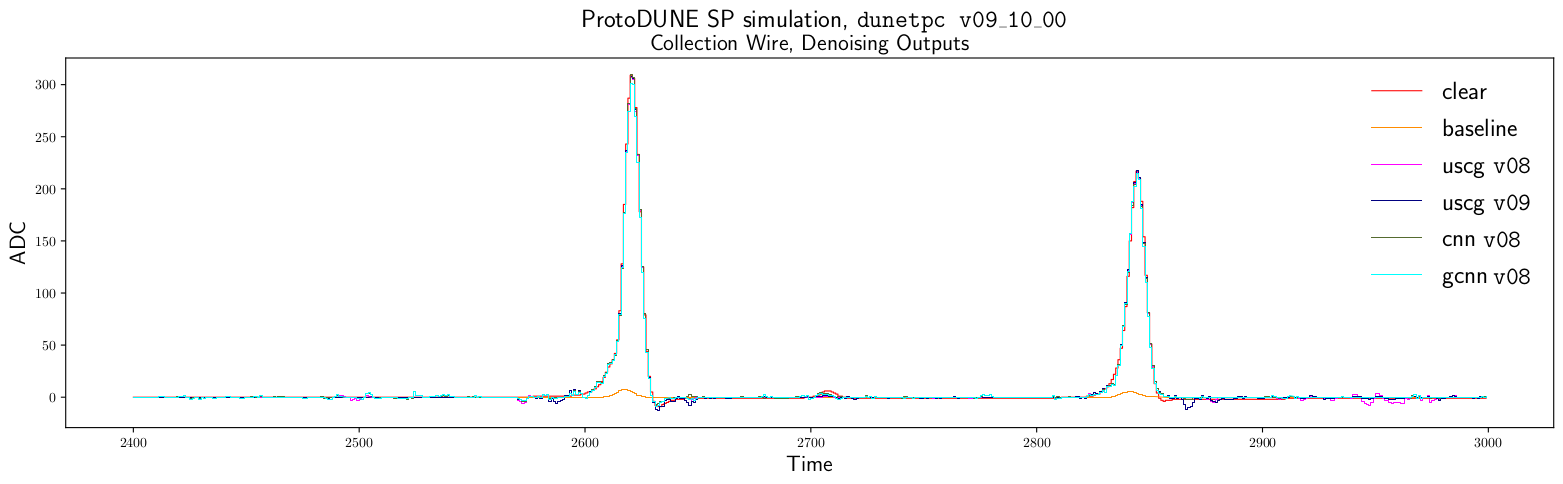}
  \caption{Detail of a raw waveform from  \dunetpc \vnine dataset: label,
  traditional algorithm and neural networks outputs. The version,
  \texttt{v08} or \texttt{v09}, next to the model name in the legend
  refers to which dataset the correspondent model was trained on.}
  \label{fig:out v9}
\end{figure*}

PSNR between a noise-free image $\mathbf{x}$ and a denoised $\mathbf{y}$ one is
given by the following equation:
\begin{equation}
  \mathrm{PSNR}(\mathbf{x}, \mathbf{y}) = 10\cdot \log_{10} \bigg(\frac{\max(\mathbf{x})^2}{\mathrm{MSE}(\mathbf{x},\mathbf{y})}\bigg)
\end{equation}
Note that PSNR and stat-SSIM increase with the reconstruction quality, while MSE
decreases. We observe that these three quantites are really suited to compare the deep
learning models between themselves, however they are not really informative when
the baseline approach is considered.
Indeed, the baseline approach aims at fitting gaussian peaks in regions
marked as interesting, rather than reconstructing the precise shape of the spike
contained in the raw digits. Hence the baseline tool performs
inevitably poorly on the considered metrics and up to our knowledge, there is no
default metric to assess its performance. Nonetheless, we define a custom
quantity that tries to compare the different approaches, observing that the
deconvolution process does not preserve waveform amplitudes, but their integrals.
For such reason we decide to evaluate the mean absolute error on wires
integrated charge (iMAE):
\begin{equation}
  \label{eq:imae}
  \mathrm{iMAE}(\mathbf{x}, \mathbf{y}) = \frac{1}{n_w}\sum_{w=1}^{n_{w}} \Big|\sum_t (\mathbf{x} - \mathbf{y})_{wt} \Big|
\end{equation}
where the sum inside the absolute values runs over time for the whole readout window,
while the outer sum averages the result over the wire dimension.
Note that the deconvolution approach does not preserve waveform amplitudes, because a
filtering function is applied to the signal in Fourier space, then results
are deconvolved back to the time domain. Furthermore, the deconvolution outputs are
known up to an overall normalization constant, that we fit on the datasets in
order to minimize the iMAE quantity. We show that, although we perform this
operation on the state-of-the-art tool outputs for a fair comparison against
our models, they nonetheless achieve a worse iMAE score. Table~\ref{tab:res-v08}
collects the metrics values evaluated on \veight dataset. We gather \vnine
dataset results in table~\ref{tab:res-v09}. Note that we present only evaluations
for $\SI{2}{\GeV}$ beam energy events, since we find metrics distributions to be
flat in the energy parameter. Figures~\ref{fig:out v8} and~\ref{fig:out v9} show
samples of labels and denoised waveforms.

USCG-Net like networks exceed GCNN-like ones in all the collected metrics. In
order to have a first assessment of the quality of the neural network generalization power, we train
two versions of the USCG-Net: one on the \veight dataset and the other on the \vnine
dataset. We decide not to train the GCNN-like networks on the \vnine dataset after
we observed difficulties in training convergence as well as long training times
on such a big dataset. We evaluate these networks on both datasets.

Following expectations, the networks trained and applied on the same dataset lead
to better performance. The only exceptions are given by the iMAE columns,
where the USCG-Net trained on the dataset opposite to the testing one, achieve the best iMAE score.
The stat-SSIM index score drops significantly for GCNN-like networks.
All the networks, nonetheless, show hints of overall good generalization power when they are
applied on datasets not used for training. This fact is well supported by the PSNR
columns, which show that even the worst model achieves competitive results. We underline
that the USCG-Net is not trained according to the stat-SSIM quantity: adding an extra 
term in the loss function containing such term could be considered a point of
further development of the present research.

\setlength\tabcolsep{2pt}
\begin{table}
  \caption{
      Test metrics for denoising on \veight dataset. Results are shown for
      collection plane and $\SI{2}{\GeV}$ beam energy only. The version,
      \texttt{v08} or \texttt{v09}, next to the model name in the first column
      refers to which dataset the correspondent model was trained on.
  }
  \label{tab:res-v08}
  \scriptsize
  \begin{tabular}{lllll}
    \hline\noalign{\smallskip}
    Model                      & stat-SSIM                 & PSNR                & MSE                & iMAE         \\
    \noalign{\smallskip}\hline\noalign{\smallskip}
    Baseline & \multicolumn{1}{c}{-}& \multicolumn{1}{c}{-} & \multicolumn{1}{c}{-} & $5391\pm 1622$    \\
    CNN v08  & $0.471\pm 0.008$       & $67.3\pm 1.2$         & $0.57\pm 0.03$        & $287\pm 12$       \\
    GCNN v08 & $0.512\pm 0.011$     & $70.12\pm 1.4$        & $0.30\pm 0.01$        & $191.4\pm 2.6$    \\
    USCG v08 & $\bf 0.988\pm 0.005$ & $\bf 72.66\pm 1.54$   & $\bf 0.17\pm 0.02$    & $95.5\pm 8.5$     \\
    USCG v09 & $0.926\pm 0.007$     & $72.3\pm 1.5$         & $0.18\pm 0.02$        & $\bf 76.3\pm 8.2$ \\
    \hline\noalign{\smallskip}
  \end{tabular}
\end{table}

\setlength\tabcolsep{2.4pt}
\begin{table}
  \caption{
        Test metrics for denoising on \vnine dataset. Results are shown for
        collection plane and $\SI{2}{\GeV}$ beam energy only. The version,
        \texttt{v08} or \texttt{v09}, next to the model name in the first column
        refers to which dataset the correspondent model was trained on.
    }
  \label{tab:res-v09}
  \scriptsize
    \begin{tabular}{lllll}
    \hline\noalign{\smallskip}
    Model                      & stat-SSIM            & PSNR              & MSE                & iMAE $[\times 10^3]$ \\
    \noalign{\smallskip}\hline\noalign{\smallskip}
    Baseline & \multicolumn{1}{c}{-}& \multicolumn{1}{c}{-} & \multicolumn{1}{c}{-} & $5.86\pm 0.52$       \\
    CNN v08  & $0.37\pm 0.02$     & $57.3\pm 1.4$         & $5.79\pm 0.88$        & $4.16 \pm 0.36$      \\
    GCNN v08 & $0.40 \pm 0.02$    & $57.7\pm 1.5$         & $5.27\pm 0.69$        & $4.51 \pm 0.39$      \\
    USCG v08 & $0.65\pm 0.05$     & $61.1\pm 1.6$         & $2.3\pm 0.2$          & $\bf 2.18\pm 0.29$       \\
    USCG v09 & $\bf 0.81\pm 0.07$ & $\bf 61.8\pm 1.7$     & $\bf 1.99\pm 0.19$    & $2.25\pm 0.23$   \\
    \noalign{\smallskip}\hline
  \end{tabular}
\end{table}

\section{Conclusions}
\label{sec:concl}
In this paper we presented Deep Learning strategies for denoising raw digits
simulated data at ProtoDUNE SP. Our approach leads to an automated tool that
cleans raw inputs and in the future could be adapted to work on real data. We
investigated the capabilities of Graph Neural Networks, testing approaches
alternative to classical Convolutional Neural Networks on a novel use case.
Graph Neural Networks aimed to exploit long distance correlations between pixels,
in particular the USCG-Net exemplified this approach, processing
big sized inputs while still fitting GPU memory constraints. Our trained neural
networks were able to outperform the state-of-the-art traditional reconstruction
\linebreak algorithm on the custom iMAE metric. All the networks gave hints of interesting generalization power, especially in the
PSNR quality assessment, achieving high performances even on datasets different
to the one they had been trained on. Further work directions will aim to fully assess
the models on larger, more complete testing datasets.

\begin{acknowledgements}
  We thank the IBM Company that \linebreak supported the realization of this paper, gently
  providing hardware and intellectual contribution.
\end{acknowledgements}

%
\section*{Conflict of interest}
The authors declare that they have no conflict of interest.

\section*{Data Availability Statement}
This manuscript has associated data in a data repository. [Authors' comment: No
associated data except for code. The associated code to replicate the studies in
this paper can be found at:
\href{https://zenodo.org/record/5821522}{\tt this URL}.

\bibliographystyle{spphys}       
\bibliography{../blbl}

\begin{thebibliography}{10}
\providecommand{\url}[1]{{#1}}
\providecommand{\urlprefix}{URL }
\expandafter\ifx\csname urlstyle\endcsname\relax
  \providecommand{\doi}[1]{DOI \discretionary{}{}{}#1}\else
  \providecommand{\doi}{DOI \discretionary{}{}{}\begingroup
  \urlstyle{rm}\Url}\fi

\bibitem{domine2020}
L.~Domin\'e, K.~Terao, Phys. Rev. D \textbf{102}, 012005 (2020).
\newblock \doi{10.1103/PhysRevD.102.012005}.
\newblock \urlprefix\url{https://link.aps.org/doi/10.1103/PhysRevD.102.012005}

\bibitem{abi2020cnn}
B.~Abi, R.~Acciarri, et~al., Phys. Rev. D \textbf{102}, 092003 (2020).
\newblock \doi{10.1103/PhysRevD.102.092003}.
\newblock \urlprefix\url{https://link.aps.org/doi/10.1103/PhysRevD.102.092003}

\bibitem{aurisano2016}
A.~Aurisano, A.~Radovic, et~al., Journal of Instrumentation \textbf{11}(09),
  P09001–P09001 (2016).
\newblock \doi{10.1088/1748-0221/11/09/p09001}.
\newblock \urlprefix\url{http://dx.doi.org/10.1088/1748-0221/11/09/P09001}

\bibitem{kronmueller2019}
M.~Kronmueller, T.~Glauch.
\newblock Application of deep neural networks to event type classification in
  icecube.
\newblock \url{https://arxiv.org/abs/1908.08763} (2019)

\bibitem{albertsson2019}
K.~Albertsson, P.~Altoe, et~al.
\newblock Machine learning in high energy physics community white paper.
\newblock \url{https://arxiv.org/abs/1807.02876} (2019)

\bibitem{bourilkov2019}
D.~Bourilkov, International Journal of Modern Physics A \textbf{34}(35),
  1930019 (2019).
\newblock \doi{10.1142/s0217751x19300199}.
\newblock \urlprefix\url{http://dx.doi.org/10.1142/S0217751X19300199}

\bibitem{abi2020tdrI}
B.~Abi, R.~Acciarri, et~al., JINST \textbf{15}(08), T08008 (2020).
\newblock \doi{10.1088/1748-0221/15/08/T08008}.
\newblock \urlprefix\url{https://doi.org/10.1088/1748-0221/15/08/T08008}

\bibitem{abi2020tdrII}
B.~Abi, R.~Acciarri, et~al.
\newblock Deep underground neutrino experiment (dune), far detector technical
  design report, volume ii: Dune physics.
\newblock \url{https://arxiv.org/abs/2002.03005} (2020)

\bibitem{abi2020tdrIII}
B.~Abi, R.~Acciarri, et~al., JINST \textbf{15}(08), T08009 (2020).
\newblock \doi{10.1088/1748-0221/15/08/T08009}.
\newblock \urlprefix\url{https://doi.org/10.1088/1748-0221/15/08/T08009}

\bibitem{abi2020tdrIV}
B.~Abi, R.~Acciarri, et~al., JINST \textbf{15}(08), T08010 (2020).
\newblock \doi{10.1088/1748-0221/15/08/T08010}.
\newblock \urlprefix\url{https://doi.org/10.1088/1748-0221/15/08/T08010}

\bibitem{abi2017pdunetdr}
B.~Abi, R.~Acciarri, et~al.
\newblock {The Single-Phase ProtoDUNE Technical Design Report}.
\newblock \url{https://arxiv.org/abs/1706.07081} (2017)

\bibitem{church2014larsoft}
E.D. Church.
\newblock Larsoft: A software package for liquid argon time projection drift
  chambers.
\newblock \url{https://arxiv.org/abs/1311.6774} (2014)

\bibitem{acciarri2017}
R.~Acciarri, C.~Adams, et~al., Journal of Instrumentation \textbf{12}(08),
  P08003 (2017).
\newblock \doi{10.1088/1748-0221/12/08/p08003}.
\newblock \urlprefix\url{http://dx.doi.org/10.1088/1748-0221/12/08/P08003}

\bibitem{adams2018}
C.~Adams, R.~An, et~al., Journal of Instrumentation \textbf{13}(07), P07006
  (2018).
\newblock \doi{10.1088/1748-0221/13/07/p07006}.
\newblock \urlprefix\url{https://doi.org/10.1088/1748-0221/13/07/p07006}

\bibitem{valsesia2019deep}
D.~Valsesia, G.~Fracastoro, et~al.
\newblock Deep graph-convolutional image denoising.
\newblock \url{https://arxiv.org/abs/1907.08448} (2019)

\bibitem{valsesia2019image}
D.~Valsesia, G.~Fracastoro, et~al.
\newblock Image denoising with graph-convolutional neural networks.
\newblock \url{https://arxiv.org/abs/1905.12281} (2019)

\bibitem{simonovsky2017dynamic}
M.~Simonovsky, N.~Komodakis.
\newblock Dynamic edge-conditioned filters in convolutional neural networks on
  graphs.
\newblock \url{https://arxiv.org/abs/1704.02901} (2017)

\bibitem{xie2017aggregated}
S.~Xie, R.~Girshick, others, P.~Dollár, Z.~Tu, K.~He.
\newblock Aggregated residual transformations for deep neural networks.
\newblock \url{https://arxiv.org/abs/1611.05431} (2017)

\bibitem{liu2020scgnet}
Q.~Liu, M.~Kampffmeyer, et~al.
\newblock Scg-net: Self-constructing graph neural networks for semantic
  segmentation.
\newblock \url{https://arxiv.org/abs/2009.01599} (2020)

\bibitem{ronneberger2015unet}
O.~Ronneberger, P.~Fischer, et~al.
\newblock U-net: Convolutional networks for biomedical image segmentation.
\newblock \url{https://arxiv.org/abs/1505.04597} (2015)

\bibitem{scarselli:2009}
F.~Scarselli, M.~Gori, et~al., IEEE Transactions on Neural Networks
  \textbf{20}(1), 61 (2009).
\newblock \doi{10.1109/TNN.2008.2005605}.
\newblock \urlprefix\url{https://doi.org/10.1109/TNN.2008.2005605}

\bibitem{xu2019powerful}
K.~Xu, W.~Hu, et~al.
\newblock How powerful are graph neural networks?
\newblock \url{https://arxiv.org/abs/1810.00826} (2019)

\bibitem{j.2018on}
S.J. Reddi, S.~Kale, et~al., in \emph{International Conference on Learning
  Representations} (2018).
\newblock \urlprefix\url{https://openreview.net/forum?id=ryQu7f-RZ}

\bibitem{kingma2017adam}
D.P. Kingma, J.~Ba.
\newblock Adam: A method for stochastic optimization.
\newblock \url{https://arxiv.org/abs/1412.6980} (2017)

\bibitem{chann:2008}
S.S. Channappayya, A.C. Bovik, et~al., IEEE Transactions on Image Processing
  \textbf{17}(6), 857 (2008).
\newblock \doi{10.1109/TIP.2008.921328}.
\newblock \urlprefix\url{https://doi.org/10.1109/TIP.2008.921328}

\bibitem{wang:2009mse}
Z.~Wang, A.C. Bovik, IEEE Signal Processing Magazine \textbf{26}(1), 98 (2009).
\newblock \doi{10.1109/MSP.2008.930649}.
\newblock \urlprefix\url{https://doi.org/10.1109/MSP.2008.930649}

\bibitem{zhao2018loss}
H.~Zhao, O.~Gallo, et~al.
\newblock Loss functions for neural networks for image processing.
\newblock \url{https://arxiv.org/abs/1511.08861} (2018)

\end{thebibliography}

\end{document}